\documentclass{article}
\usepackage{epsfig}
\usepackage[dvips]{color}
\usepackage{colordvi}  

\begin{document}

%
%%%%%%%%%%%%%%%%%%%%%%%%%%%%%%%%%%%%%%%%%%%%%%%%%%%%%%%%%%%%%%%%%%%%%%%%%%
%

\begin{center}
\Large{\bf Rescattering contributions to final state interactions
in (e,e'p) reactions\footnote{\small To appear in the proceedings of
 "2$^{\rm nd}$ International Conference on Nuclear and Particle Physics
  with CEBAF at  JLab."}}
\end{center}

\bigskip

\begin{center}
\large  C.~Barbieri${}^{a}$\footnote{\small Email: \tt barbieri@triumf.ca},
% W.~H.~Dickhoff${}^{b}$,
 L.~Lapik\'as${}^{b}$,
%  H.~M\"uther${}^{d}$,
   D.~Rohe${}^{c}$
\end{center}

\medskip

\begin{center}
${}^{a}$TRIUMF,  Vancouver, B.C., CANADA V6T 2A3 \\
${}^{b}$NIKHEF,  1109DB Amsterdam, The Netherlands \\
${}^{c}$Department f\"ur Physik und Astronomie,
                  Universit\"at Basel, CH-4056 Basel, Switzerland   \\ 
\end{center}

\medskip

\noindent
 A semiclassical model is employed to study the effects of 
rescattering on $(e,e'p)$ cross sections. We consider a two-step process
with the propagation of an intermediate nucleon and use Glauber theory
to account for the effects of N--N scattering.
 This calculation has relevance for the analysis of data
at high missing energies. Of particular
interest is the E97-006 experiment done at JLab.
 It is found that rescattering is strongly reduced in parallel
kinematics and that the excitation of nucleon resonances
is likely to give important contributions to the final-state interactions
in the correlated region.

\bigskip

%\preprint{TRI-PP-03-29}
\noindent
TRIUMF preprint: TRI-PP-03-29



%%%%%%%%%%%%%%%%%%%%%%%%%%%%%%%%%%%%%%%%%%%%%%%%%%%%%%%%%%%%%%%%%%%%%%%%%%

\section{Introduction and motivations}

 Short-range correlations strongly influence the dynamics
of nuclear systems. The repulsive core at small 
internucleon distances has the effect of removing the nucleons from
their shell model orbitals, producing pairs of nucleons with high
and opposite relative momenta. 
 This results in spreading out a sizable amount of spectral strength, 
about 10-15\%~\cite{bv91}, to very high missing energies and momenta and
in increasing the binding energy of the system.
 Theoretical studies of the distribution of short-range correlated 
nucleons for finite nuclei have been carried out in Ref.~\cite{WHA}
and by Benhar et al.~\cite{Benhar}. These calculations suggest that most
of this strength is found along a ridge in the momentum-energy ($k$-$E$) 
plane that spans several hundreds of MeV/c (and MeV). The most probable
energy is the one of a free moving nucleon but shifted by a constant
term that represents the two-hole potential well in which the
correlated pair moves.
 Evidence of two nucleon structures, similar to the deuteron, inside
finite nuclei has also been discussed using the variational approach
in Ref.~\cite{SchiavillaSR} and found to be driven by
short-range and tensor correlations.
We note that short-range effects alone do not completely explain
the depletion of single particle orbitals near the
Fermi energy~\cite{Louk93},
which require a proper description of low-energy collective
modes~\cite{LRC}.
However, a proper understanding of short-range effects
remains of great importance for the understanding of nuclear
structure~\cite{Wim03}.

 \begin{figure}[t]
\centerline{\psfig{figure=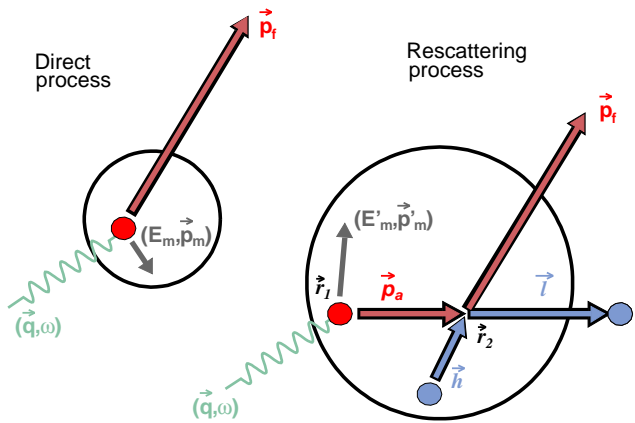,width=9cm,angle=0}}
\caption{ Schematic representation of the direct knockout of a proton,
  given by the PWIA (left) and the contribution from a two-step
  rescattering (right). In the latter a proton or neutron is emitted
  with momentum $\vec{p}_a$ and different missing energy and momentum
  $(E_m',\vec{p}_m')$. Due to a successive collision, a proton
  is finally detected with the same momentum $\vec{p}_f$ seen in the
  direct process.
  }
\label{fig:TotRes}
\end{figure}

Experimentally, $(e,e'p)$ reactions have been used for  a long
time to determine spectral functions at small missing energy. 
Past measurements in the region of interest to short-range
correlations have been limited due to the enormous background
that is generated  by final state interactions (FSI).
The issue of how to minimize the FSI has been addressed
in Ref.~\cite{E97proposal}. There, it is shown that exclusive $(e,e'p)$
cross sections are dominated by two-step processes like the one
depicted in Fig.~\ref{fig:TotRes}. This becomes particularly 
relevant when perpendicular kinematics are employed to probe
the regions of small spectral strength.
 In Ref.~\cite{E97proposal}, it was suggested that the 
contribution of rescattering can be diminished in parallel kinematics.
 New data was subsequently taken
in these conditions at Jefferson Lab~\cite{danielaElba}.

The theoretical calculation of the rescattering yield has been
addressed in Ref.~\cite{Marcel}. This calculation was based on the
semiclassical model of Ref.~\cite{VjPi} and intended to be 
applied at lower missing energies. This contribution reports
about an ongoing work aimed to extend these calculations
to the kinematics of interest for the study of short-range
correlations.

%%%%%%%%%%%%%%%%%%%%%%%%%%%%%%%%%%%%%%%%%%%%%%%%%%%%%%%%%%%%%%%%%%%%%%%%%%%%%%%

\section{Model}

At large $E_m$ appreciable contributions to the experimental yield come
from two-step
mechanisms, in which a reaction $(e,e'a)$ is followed by a scattering process
from a nucleon in the nuclear medium, $N'(a,p)N''$, eventually leading
to the
emission of the detected proton. In general, $a$ may represent
a nucleon or another possible intermediate particle, as a $\Delta$
excitation. In the
following we will also use the letter $a$ to label the different open channels.

Following the semiclassical approach proposed in Refs.~\cite{VjPi,Marcel},
we write  the contribution to the cross section coming from rescattering
through the channel $a$ as
\begin{eqnarray}
    { d^6 ~ \sigma^{(a)}_{rescat}
     \over
     dE_0 \; d\Omega_{\hat{k}_o}  dE_f \; d \Omega_{\hat{p}_f} } &=&  
    \int_V d \vec{r}_1 \int_V d \vec{r}_2 \int_{0}^{\omega} d T_a
  \rho_N(\vec{r}_1) \; 
    { K \; S^h_N(p_m,E_m) \; \sigma^{cc1}_{eN}
    \over
    A \; ( \vec{r}_1 - \vec{r}_2 )^2 } \; \; \;
\nonumber  \\
 &\;& \times 
 P_T(p_a, \vec{r}_1 , \vec{r}_2 )
\rho_{N'}(\vec{r}_2) \; 
    { d^3 ~ \sigma_{a N'}
    \over
    dE_f \; d \Omega_{\hat{p}_f} } \;
P_T(p_f, \vec{r}_2 , \infty) \; ,
\label{eq:TotRes}
\end{eqnarray}
where $(E_o,\vec{k}_o)$ and  $(E_f,\vec{p}_f)$ represent the four-momenta
of the outgoing electron and proton,  respectively.
Eq.~(\ref{eq:TotRes}), assumes that the intermediate particle $a$ is
generated in PWIA by the electromagnetic current at a point $\vec{r}_1$
inside the nucleus. Here $K=|\vec{p}_a|E_a$ is a phase space factor,
$S^h_N(p_m,E_m)$
the spectral function of the hit nucleon $a$ and $\sigma^{cc1}_{eN}$
is the off shell electron-nucleon cross section~\cite{deForest}.
The transparency factor $P_T(p, \vec{r}_1 , \vec{r}_2)$ gives
the probability that the struck nucleon $a$ propagates to
a second point $\vec{r}_2$, where it scatters from the nucleon $N'$ with
cross section $d^3 ~ \sigma_{a N'}$. The whole process is depicted
in Fig.~\ref{fig:TotRes}.

In the calculation described in Sec.~\ref{sec:results} we will only
consider the channels in which $a$ is either a proton or a neutron.
It is clear that other channels are expected to be important.
In particular, the excitation of the $\Delta$ resonance is also seen 
to contribute from the preliminary data of the E97-006
experiment~\cite{danielaElba}.

 Eq.~(\ref{eq:TotRes}) is a seven-fold integral that can be conveniently 
evaluated with Monte Carlo techniques, once the terms in the integrand
are known.
In the following we describe  the calculation of the cross section
$d^3 ~ \sigma_{a N'}$ and of the transmission probability $P_T$.

\subsection{Evaluation of the in-medium nucleon-nucleon rate}

For the present purposes the spectral distribution of the hit nucleon,
$N'$, can
be appropriately described by the free Fermi gas distribution.
The cross section is therefore computed for a nucleon $a$ travelling
in symmetric nuclear matter at a given density $\rho_{NM}$.
The nucleon $N'$ must have
a momentum $\vec{h}$ smaller than the Fermi momentum
$k_f = (3\pi^2\rho_{NM}/2)^{1/3}$.
 Among all the nucleons involved in the process,
$\vec{p}_f$ will refer to the outgoing proton while the others can be
either neutrons or protons depending on the channel $a$.
 The finite size effects are eventually included in Eq.~(\ref{eq:TotRes})
using the local density approximation, that is, by evalutaing the
cross section for the density at the point $\vec{r}_2$.

The probability per unit time of an event leading to the emission
of a proton with momentum $\vec{p}_f$ is obtained by imposing the
Pauli constraints and integrating over the unobserved
momenta $\vec{h}$ and $\vec{l}$. Employing a relativistic
notation,
\begin{eqnarray}
d^3 P 
%(p_a , \vartheta_{af}, \rho_{NM}) 
 \over dp_f d\Omega_{\hat{p}_f} &=&
  2 \; \vartheta(p_f - k_f)  \; L^3
      \int  \int_{L^3} {d\vec{h} \;  d\vec{l} \over (2\pi)^6} 
      \; \vartheta(k_f - h) \vartheta(l - k_f) \;
         W_I
\nonumber \\
 &=& 2 \; p_f^2 \; \vartheta(p_f - k_f) \int_{L^3} 
                  {d\vec{h} \over (2\pi)^3} 
    \vartheta(k_f - h) \vartheta(l - k_f)
            { m_a \; m_h \over E_a(p_a) E_h(h)} 
\label{eq:P_dpf}  \\
  & & ~ ~ ~\left. 
   \times { {\cal M}(s,t,u) \over  4 \pi^2}
    { m_p \; m_l \over E_p(p_f) E_l(l)} \delta(E_a + E_h - E_f - E_l)
 \right|_{\vec{l} = \vec{p}_a + \vec{h}_f - \vec{p}_f} \; ,
\nonumber 
\end{eqnarray}
where $L^3$ is the volume of a normalization box
and $E_N(p) = (p^2 + m_N^2)^{1/2}$.
In Eq.~(\ref{eq:P_dpf}), $W_I$ is the probability per unit time for
the event $p_a^\mu + h^\mu \rightarrow p_f^\mu + l^\mu$  which is
expressed in terms of the Lorentz 
invariant amplitude ${\cal M}(s,t,u)$~\cite{PeSch}.
 In the present work, we use the in vacuum
values for ${\cal M}(s,t,u)$  and extract them from the SAID
phase shift data analysis~\cite{SAID}.

The in medium scattering rate is finally related to Eq.~(\ref{eq:P_dpf}) by
\begin{equation}
  { d^3 ~ \sigma_{a N'}
    \over
    dE_f \; d \Omega_{\hat{p}_f} }
    = {E_a \over \rho_{N'} p_a }{ E_f \over p_f}
  { d^3 P   \over dp_f d\Omega_{\hat{p}_f} }
    \; .
 \label{eq:SaN_dpf}
\end{equation}

\subsection{Transparency factor}

 According to Glauber theory,  the
probability $P_T$ that a proton struck at $\vec{r}_1$ will travel
with momentum $\vec{p}$ to the point $\vec{r}_2$ without being
rescattered is given by
\begin{eqnarray}
P_T(p,\vec{r}_1,\vec{r}_2) = & 
exp \left\{ - \int_{z_1}^{z_2}  \right.dz & 
    \left[ g_{pp}(|\vec{r}_1 - \vec{r}|) \; \tilde{\sigma}_{pp}(p,\rho(\vec{r}))
           \; \rho_p(\vec{r})   \right.
\label{eq:PT} \\
   & &  \left. \left. ~+~  g_{pn}(|\vec{r}_1 - \vec{r}|) \; \tilde{\sigma}_{pn}(p,\rho(\vec{r}))
           \; \rho_n(\vec{r}) \right]  \right\} \; ,
\nonumber
\end{eqnarray}
where the z axis is chosen along the direction of propagation
$\vec{p}$, an impact
parameter $\vec{b}$ is defined so that $\vec{r} = \vec{b} + z \hat{p}$,
and $z_1$ ($z_2$) refer to the initial (final) position.
The in medium total cross sections $\tilde{\sigma}_{pp}(p,\rho)$ and
$\tilde{\sigma}_{pn}(p,\rho)$ have been computed in Ref.~\cite{VjPi}
up to energies of 300~MeV and account for the effects of Pauli blocking,
Fermi spreading and the velocity dependence of the nuclear mean field.
 For energies above 300~MeV these
have been extended to incorporate effects of pion emission~\cite{Pieperpriv}.
The pair distribution functions
$g_{pN}(|\vec{r} - \vec{r}_1|)$ account for the joint probability of 
finding a nucleon N in $\vec{r}$ and a proton at $\vec{r}_1$.

The nuclear transparency is defined, in Glauber theory,
as the average over the nucleus of the probability that the struck proton
emerges from the nucleus without any collision. This is related
to $P_T$ by
\begin{equation}
 T = {1 \over Z} \int d\vec{r} \rho_p(\vec{r}) P_T(p,\vec{r},\infty) \; .
 \label{eq:T}
\end{equation}
For the nucleus of ${}^{12}{\rm C}$ and an outgoing proton of
energy $E_f \sim$~1.8~GeV, which is of interest for the present
application, we find $T =$~0.62.

%%%%%%%%%%%%%%%%%%%%%%%%%%%%%%%%%%%%%%%%%%%%%%%%%%%%%%%%%%%%%%%%%%%%%%%%%%%%%%

\section{Results}
\label{sec:results}

\begin{figure}[t]
\centerline{\psfig{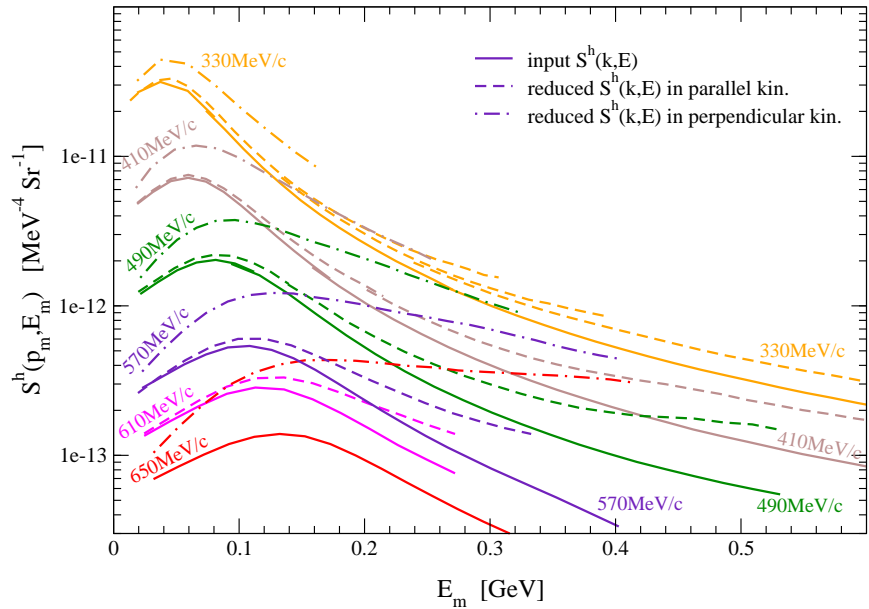}}
\caption{ 
  Theoretical results for the reduced spectral strength in the correlated
 region obtained in parallel (dashed line) and perpendicular
 (dot-dashed line) kinematics.  The full line shows the model spectral
 function, Eq.~(\ref{eq:Shtotal}), employed in the calculations.
 }
\label{fig:res1}
\end{figure}

At energies close to the Fermi level the hole spectral function 
is dominated by contribution from the mean field orbitals
in $s$ and $p$ shells.
These are known from ${}^{12}{\rm C}(e,e'p)$
experiments and describe about 60~\% of the total
distribution~\cite{LoukC12}.
 It is therefore convenient to split the spectral function in a mean field
and a correlated part, 
\begin{equation}
 S^h_p(p_m,E_m) = S^h_{MF}(p_m,E_m) + S^h_{corr}(p_m,E_m) \; ,
 \label{eq:Shtotal}
\end{equation}
in which $S^h_{corr}(p_m,E_m)$ also contains the short-range correlated
tail at very high missing energies and momenta~\cite{WHA,Benhar}. 
In the following we parametrize this as
\begin{equation}
 S^h_{corr}(p_m,E_m) = 
   { C \;  e^{- \alpha \, p_m} \over [E_m - e(p_m)]^2 + [\Gamma(p_m)/2]}
\end{equation}
where $e(p_m)$ and $\Gamma(p_m)$ are smooth functions and all
parameters were chosen to give an appropriate fit to the available data
in parallel kinematics~\cite{danielaElba}.
The solid line in Fig.~\ref{fig:res1} shows the model spectral function
(\ref{eq:Shtotal}) employed in the calculation in that part of the 
$k$-$E$ plane where $S_{corr}$ dominates.

We have performed calculations of the rescattering contribution by
employing both parallel and perpendicular kinematics.
 In the first case,
the angle  between the momenum transfered by the 
electron and the momentum of the final proton was chosen to be
$\vartheta_{qf} \sim$~5~deg
and the energy of the final proton was $E_f\sim$~1.6~GeV.
 For the perpendicular kinematics, $\vartheta_{qf}\sim$~30~deg
and $E_f\sim$~1~GeV. In both cases the four momentum transfered by
the electron was $Q^2 \sim$~0.40~GeV$^2$.

The results for the total cross section 
($\sigma_{PWIA}+\sigma_{rescat}$) have been converted to a reduced spectral 
function representation by dividing them by $|p_f E_f|T\sigma_{eN}^{cc1}$,
evaluated for the kinematics of the direct process (see Fig.~\ref{fig:res1}).
 As can be seen, FSI from nucleon-nucleon rescattering give little
contribution to the total cross section in parallel kinematics, and
the resulting reduced spectral function is close to the true one.
For perpendicular kinematics, more sizable contributions are found
and they tend to fill the region at higher missing energies,
where the spectral function is small.
 This confirms the trend of FSI expected for parallel kinematics
that strength primarly is moved from places, where  $S^h(p_m,E_m)$
is small to places, where it is large, and thus gives a small relative
effect~\cite{E97proposal}.
To check that the contribution found  is actually coming from the 
correlated part of the spectral function itself,
we have repeated the calculation in perpendicular kinematics
by neglecting $S_{corr}$ in
Eq.~(\ref{eq:Shtotal}). For missing momenta above 400~MeV no
rescattering from nucleons was found in the mean field region.
One should note that since little reliable experimental information for $S_{corr}$
is available to date, the correlated strength can be extracted from the
experimental data only in a self-consistent fashion.
This of course requires a proper treatment of the FSIs.

\begin{figure}[t]
%\vspace{.2in}
\centerline{\psfig{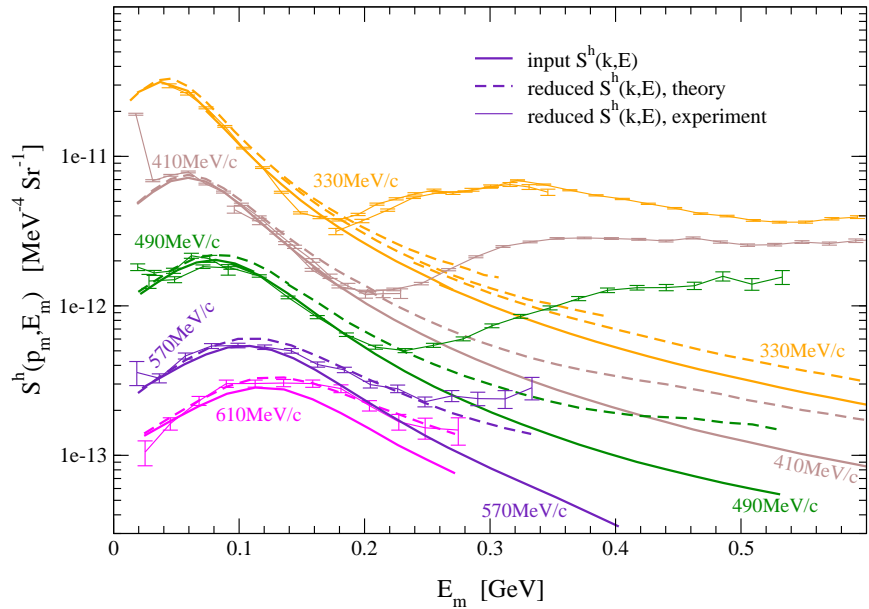}}
\caption{ 
  Theoretical results for the reduced spectral strength obtained in
  parallel kinematics (dashed line) compared to the experimental
  results of Ref.~\cite{danielaElba}.
    The full line shows the model spectral function of Eq.~(\ref{eq:Shtotal})
  employed in the calculations.
}
\label{fig:res2}
\end{figure}

Fig.~\ref{fig:res2} compares the model spectral function~(\ref{eq:Shtotal})
and the theoretical reduced one, with preliminary results from
the E97-006 collaboration~\cite{danielaElba}. An enhancement of the
cross section is found esperimentally at very high missing energies that
is presumably generated by the excitation of a $\Delta$ resonance.
 This effect
is not included in the present calculation yet. Contributions
from rescattering through this channel are expected to fill up the valley
between the correlated and $\Delta$ regions more substantially for heavier
nuclei.
 Therefore, these additional degrees of freedom need to be included
in the present model.

%%%%%%%%%%%%%%%%%%%%%%%%%%%%%%%%%%%%%%%%%%%%%%%%%%%%%%%%%%%%%%%%%%%%%%%%%%%%%%%

\section{Conclusions}

 This contribution reports about an ongoing work aimed to study the
effects of final state interactions in $(e,e'p)$ reactions, as
generated by rescattering effects.
 The two-step rescattering processes that involve the intermediate
propagation of a nucleon have been approached by using a semiclassical
model.
A preliminary calculation has been reported for ${}^{12}{\rm C}(e,e'p)$.
It is found that the contribution from final state interactions
in parallel kinematics is much smaller than in perpendicular ones.
In the latter case a large
amount of strength is shifted from regions where the spectral function
is big to regions where it is smaller, thus overwhelming the experimental
yield from the direct process.
This confirms the studies of Ref.~\cite{E97proposal}.

At $E_m >250$MeV the present experimental results exceed the calculated direct 
plus rescattering contributions by about an order of magnitude. This is 
presumably due to the excitation of $\Delta$ resonances. It suggests that 
rescattering 
through this channel also affects the measurements in the correlated region.
 The inclusion of these effects in the present model 
will be the topic of future work.

\begin{center}
{\bf Acknowledgements}
\end{center}

One of us (C.B.) would like to acknowledge a useful discussion with
Prof.~I.~Sick and the hospitality of the
Institut f\"ur Theoretische Physik at the University of T\"ubingen,
where this work was started.
This work is supported by the Natural
Sciences and Engineering Research Council of Canada (NSERC).
The work of L.L. is part of the research programme of the ``Stichting voor
Fundamenteel Onderzoek der Materie (FOM)'', which is financially
supported by the ``Nederlandse Organisatie voor Wetenschappelijk
Onderzoek (NWO)''. 
The work of D.R. is supported by the ``Schweizerische Nationalfonds''.
%

%%%%%%%%%%%%%%%%%%%%%%%%%%%%%%%%%%%%%%%%%%%%%%%%%%%%%%%%%%%%%%%%%%%%%%%%%%%%%

%%%%%%%%%%%%%%%%%%%%%%%%%%%%%%%%%%%%%%%%%%%%%%%%%%%%%%%%%%%%%%%%%%%%%%%%%%%%%

\end{document}